\documentclass[letterpaper,10pt,twocolumn]{article}
\pdfoutput=1
\usepackage{times}
\usepackage{fullpage}
\usepackage{mathptmx}
\usepackage{ptmgreek}
\usepackage{amsmath}
\usepackage{amsthm}
\usepackage{amssymb}
\usepackage{amsfonts}
\usepackage{color}
\usepackage{graphicx}
\usepackage{caption}
\usepackage{subcaption}
\usepackage{mdwlist}
\usepackage{cite}
\usepackage{listings}
\usepackage{enumitem}
\usepackage{dblfloatfix}
\usepackage{comment}
\usepackage[hyphens]{url}
\usepackage[pdftex,hidelinks]{hyperref}
\usepackage{xspace}

\usepackage{enumitem}
\setlist{topsep=4pt,itemsep=2pt,parsep=2pt}
\renewcommand{\paragraph}[1]{\smallskip\noindent\textbf{#1}}

\begin{document}
\title{Cryptographically Enforced Control Flow Integrity}
\author{Ali Jos\'e Mashtizadeh, Andrea Bittau, David Mazi\`eres, Dan Boneh\\
\\
Stanford Univeristy}
\date{}

\maketitle
\thispagestyle{empty}

\thispagestyle{plain}
\pagestyle{plain}
\setcounter{page}{1}
\pagenumbering{arabic}

\begin{abstract}
Recent Pwn2Own competitions have demonstrated the continued
effectiveness of control hijacking attacks despite deployed
countermeasures including stack canaries and ASLR\@.  A powerful
defense called Control flow Integrity (CFI) offers a principled
approach to preventing such attacks.  However, prior CFI
implementations use static analysis and must limit protection to
remain practical.  These limitations have enabled attacks against all
known CFI systems, as demonstrated in recent work.

This paper presents a \emph{cryptographic} approach to control flow
integrity (CCFI) that is both fine-grain and practical:  using message
authentication codes (MAC) to protect control flow elements such as
return addresses, function pointers, and vtable pointers.  MACs on
these elements prevent even powerful attackers with random read\slash
write access to memory from tampering with program control flow.  We
implemented CCFI in Clang\slash LLVM, taking advantage of recently
available cryptographic CPU instructions.  We evaluate our system on
several large software packages (including nginx, Apache and memcache)
as well as all their dependencies.  The cost of protection ranges from
a 3--18\% decrease in request rate.



\end{abstract}

\section{Introduction}

In recent years, sophisticated attacks on software vulnerabilities
have emerged, proving that deployed protection mechanisms can be
bypassed (e.g.,\ \cite{pwn2own,heapspray,BRSS08,brop} and many
others).  The weaknesses in many deployed defenses is that they focus
on patching specific attack techniques rather than addressing
the fundamental problem.  For example, stack
canaries~\cite{stackguard} assume a stack overflow; non-executable
memory~\cite{nxbit} assumes code injection; and address space layout
randomization~\cite{aslreff} assumes that information cannot be
leaked.  These defenses can be circumvented, for example, by
overflowing the heap, executing a chain of existing code fragments
using return-oriented programming, and leaking a pointer.  A more
principled defense approach is needed.

Exploits often work by hijacking the program's control flow to
execute unintended code, for example, to start a shell.  Indeed, all
the attacks mentioned above work by hijacking control flow and all the
defenses mentioned try to prevent specific approaches to control flow 
hijacking.

A principled solution, called {\em control flow integrity} (CFI)~\cite{cfi},
prevents an attacker from arbitrarily modifying the target of indirect jumps
(e.g.,\ return addresses, function pointers).  Ensuring control flow integrity
would prevent all attacks based on control flow hijacking, which includes all
the sophisticated attacks listed above.

However, practical implementations of CFI have been shown to be 
insecure~\cite{out_of_control,stitchgadgets} for two main reasons.  First, CFI 
uses static analysis to determine the target of a pointer, which is not always 
precise and can lead to overly permissive choices.  Second, some run-time 
checks are removed or simplified to reduce performance overhead.

In practice, existing CFI systems are very coarse-grained and group function 
pointers and return addresses into two different classes, preventing swaps 
between the two.  That is, a function pointer cannot be replaced with a return 
address, however one can still swap two return addresses (or two function 
pointers).  We are the first system to enforce CFI using cryptography, and our 
system provides a fine-grain CFI implementation in attempt to avoid arbitrary 
swaps.


\paragraph{Our contribution.}
We show that on modern processors fine-grain control flow integrity can be 
efficiently achieved using cryptography.  Our cryptographic control flow 
integrity (CCFI) system identifies all objects that would affect a program's 
control flow (e.g., return addresses, function pointers) and computes a message 
authentication code (MAC) of such objects each time they are stored in memory.  
This MAC is stored along with the object and checked every time the value is 
loaded from memory.  The random secret key used for computing these MACs is 
stored in dedicated registers so that it can never leak by a memory disclosure 
bug.  By checking the MAC of every control-flow element before using it we 
prevent the attacker from writing an arbitrary memory address into, say, a 
return address.

Moreover, we include metadata in the MAC to help prevent an attacker from 
swapping any leaked function pointers. This protection is somewhat analogous to 
replay protection in cryptographic protocols, though not as complete.  We 
prevent any two pointers from being swapped by MACing location as well as the 
value of pointers from being swapped by MACing location as well as the value of
ever pointer.  Moreover, we make it less likely for two function pointers to 
pointers to reside at the same address in a predictable way by randomizing 
every memory allocation.  We cannot prevent a particular pointer from being 
replayed to an old value if the attacker knew both its old value and MAC and 
can overwrite both, but this is the least likely scenario to help an attacker.

Cryptographic CFI is a general enough to support implementing all existing CFI 
systems that we are aware of, by including the pointer class in each MAC 
computation.  Grouping pointers into classes that are represented by numerical 
tags is used by all existing systems.  There are four or fewer classes in 
almost all existing system.


To argue security we assume a very powerful adversary: one that has
arbitrary read access to all of memory and write access to all
writable control-flow elements in memory (e.g.,\ return addresses and
function pointers).  We design our system so that, with the limited
exception of same-address pointer replay, even such a powerful
adversary cannot hijack control flow and cause unintended code to run.
At most the attacker can cause the program to terminate due to a MAC
failure.

The use of MACs in CCFI gives us several useful advantages over traditional CFI 
approaches using runtime checks.  First, hackers cannot generate MACs, they 
must have previously observed a valid MAC\@.  Second, rather than relying on 
grouping pointers at compile-time, we can group pointers based on runtime 
characteristics.  For instance our implementation binds a pointer to the 
address at which it is stored, preventing an attacker from swapping pointers.  
This is critical as static analysis is hard, not practical with dynamic 
libraries, and leads to overly permissive checks.  Third, MACs are flexible 
enough to allow us to also group pointers based on compile time grouping, 
e.g.,\ type information, or return address versus function pointers.  Finally, 
our implementation uses the new AES-NI instructions to efficiently compute the 
MAC in only 40 cycles, and so checks need not be omitted or simplified for the 
sake of performance, at the cost of security.

We implemented CCFI in LLVM~\cite{llvm} for x86-64 and recompiled SPEC2006, 
three web servers (Apache, Nginx, Lighttpd), two cache servers (memcached, 
redis) and all of their 21 dependencies.  Only two packages (libapr and Nginx) 
required code changes for compatibility, where manual MAC computations had to 
be inserted, a total of three lines changed.  The implementation is 
generalizable to any architecture that supports a fast cryptographic MAC 
function.  The web servers showed a reduced request throughput rate of only 
3--18\% when serving a static file.  When running over SSL, which contends for 
AES-NI register use, the overhead is 38\%.  This is a reasonable overhead given 
the precision of CFI achieved. 
Our results show that CCFI is practical, and we argue that its level of 
security can finally put an end to control-flow based attacks.

\section{Background}

Software vulnerabilities take all shapes and forms.  The classic example is a
stack buffer overflow, where the lack of bounds checking lets an attacker
corrupt a return address on the stack causing execution to jump to an arbitrary
location, often leading to execution of arbitrary code.  Another example is
sending data past the end of a buffer over the network, possibly leaking
sensitive information.  Yet another example is forgetting to include important 
authentication steps in the program's logic.  We classify software 
vulnerabilities and attacks as follows:

\begin{enumerate}
\item \emph{Control flow attacks} result in the attacker being able to
  execute arbitrary code.  These are the most common exploits, and
  they typically yield a remote shell.

\item \emph{Data flow attacks} result in the attacker being able to
  read or write program memory, not necessarily leading to arbitrary
  code execution.  OpenSSL's heartbleed bug~\cite{heartbleed} is an
  example, where the attacker is able to read the server's private key
  from memory.

\item \emph{Logic errors} result in the attacker being able to skip
  checks.  For example, Apple's goto fail~\cite{goto_fail} bug did not
  properly check SSL certificates allowing attackers to mount
  man-in-the-middle attacks.
\end{enumerate}

Our work focuses on the first class of attacks only.  It is, however, the most
prevalent class of attack today and the most powerful, because it allows the 
attacker to achieve everything that other attack classes do.  
Running arbitrary code lets an attacker disclose memory to leak SSL
keys or jump past any checks.  Conversely, a data flow bug that merely
discloses memory (though still catastrophic) cannot be used to execute a remote
shell on the system.

The most widely deployed protections against control flow attacks are
stack canaries, non-executable memory (NX), and address space layout
randomization (ASLR).  Interestingly, none of these solutions protect
the return address itself, which is what the attacker is after.
Instead, they attempt to stop the attack indirectly through other
means.  In contrast, our CCFI directly protects return addresses and
function pointers.  

Control flow attacks generalize to exploit any indirect branch (e.g.,
function pointers) which CCFI also protects.  In
Section~\ref{sec:related} we review related work on protecting
function pointers such as PointGuard~\cite{bib:pointguard} as well as
other stack smashing defenses.

\paragraph{Control flow integrity.}
CFI~\cite{cfi} is a technique where static analysis determines where an 
indirect jump can land.  Runtime checks are added to enforce that the jump 
lands only to the valid locations that have been determined ahead of time 
during static analysis.  For example, suppose that the pointer analysis 
determines that a function pointer can only ever point to {\tt read} or {\tt 
write}.  Then the attacker cannot modify the function pointer to call {\tt 
execve} as that call is outside the set of valid locations for that particular 
function pointer.  Note, however, that the attacker can still swap a {\tt read}
for a {\tt write}, which may be enough to conduct an attack.

More fundamentally, however, static analysis has limits; there are cases
where it cannot be determined where a function pointer is allowed to point.  In
this case, the set of valid locations can be any function whose address has 
been taken.  Worse, practical CFI implementations split valid locations into 
only two sets: function pointers can jump to any function whose address has 
been taken, and return instructions can return to any return site.  These loose
implementations are not enough as an attacker can swap return sites to 
eventually execute arbitrary code and break out of 
CFI~\cite{out_of_control,stitchgadgets}.

Our approach is tackling CFI from a cryptographic point of view.  We stop the
attacker from modifying the return address or function pointer in the first
place rather than restricting its values to a particular set (which may be large
enough for an attack).  This approach does not require static analysis and so
does not inherit any of its limitations.  The pointer is secured via a MAC at 
runtime when it is first stored in memory, and validated before use.  Function 
pointer swaps are avoided by preventing replay attacks---i.e.,\ a function 
pointer is valid in only one location and cannot be reused elsewhere.  This 
lets us provide strong CFI guarantees that are not overly permissive.

\section{Threat Model}

Many security systems today (e.g., stack canaries, ASLR) assume the
attacker cannot read memory.  An attacker who can read arbitrary
memory can easily defeat these defenses as demonstrated in several
recent papers~\cite{pwn2own, brop}.

In this paper we assume a powerful attacker who has the ability to read
arbitrary areas of memory.  Moreover, the attacker can overwrite all writable
control-flow elements in memory (e.g.,\ return addresses and function pointers).
However, the attacker is unable to write to executable memory (marked read-only)
or read the value of special registers our compiler reserves.  These are
reasonable assumptions that accurately model modern control hijacking attacks.

\section{Design}
\label{sec:design}

Our Cryptographic Control Flow Integrity (CCFI) system is a compiler that
protects any known object that may affect the program's control flow.
Specifically, it protects:
\begin{itemize}
\item Return addresses and frame pointers.

\item Function pointers.

\item vtable pointers.  The vtable is a (read-only) function pointer table used
by C++.  By swapping vtable pointers, one can cause methods of a different class
to be invoked, therefore subverting control flow.

\item Exception handlers.
\end{itemize}

There are other application dependent objects which may eventually influence a
program's control flow.  For example, an integer may index a jump table, and if
overwritten, the attacker may execute a different function.  We do not offer any
automatic protection against such data-flow attacks.  The programmer can,
however, use our primitives (defined in Section~\ref{sec:primitives}) to
manually protect such sensitive objects.

CCFI protection is achieved using a MAC\@.  Each time a control-flow object is 
stored, its MAC is computed, and on each load, its MAC is verified.  The MAC is 
stored alongside the object.  Attackers cannot overwrite a control object with 
an arbitrary value because they do not posses the MAC key needed to produce a 
valid MAC for the object.  The MAC key is randomly generated at program start, 
and stored in registers the CCFI compiler reserves (In x86-64 we use 
XMM5--XMM15).  The key is never written to memory because the recompiled 
applications and libraries do not use these registers.  Attackers cannot 
execute (misaligned) code that accesses these registers because they would have 
to break control flow in the first place.  

Care must be taken to prevent replay attacks.  For example, an attacker must not
be able to swap two function pointers by reading a function pointer and its
valid MAC and using these elsewhere (in another function pointer).  To combat
replay attacks we must first define what is included in the MAC. 

\subsection{MAC Function}
\label{sec:MACfunction}

Our MAC is implemented as a single block of AES applied to the input data.
More precisely, the MAC function is defined as follows:
\begin{verbatim}
 MAC(K, pointer, class)
  AES-128(K, { pointer, class })
\end{verbatim}
Where \verb|K| is the key, and \verb|pointer| is the object being secured
and \verb|class| is the pointer class.  In all practical CFI implementations, 
both runtime and compile time, pointers are grouped into classes.  In the 
original CFI work, there were two classes function pointers and return 
pointers.  In our scheme this can easily be represented by a one or zero.  Fine 
grain CFI may group function pointers based on usage or type signature and 
create thousands of classes.  Any modern CFI system that we know about can use 
MACs instead of weak runtime checks, which can then benefit from including 
runtime groupings to offer better security.

Runtime groupings have several benefits.  Current CFI systems must determine a 
fixed set of functions within a class and these are stored as read-only tables, 
otherwise an attacker could attack the tables used to verify control flow 
integrity.  As an example CCFI can use addresses where pointers are stored to 
prevent swapping function pointers, this is not possible at compile time.  
Another benefit is that the size of runtime class can always be smaller or 
equal to the size of compile time classes.  To see why this is imagine a 
function pointer has one of five valid values, at runtime the attacker may only 
see three of them because of the program's state.  Thus the attacker can never 
exploit the remaining two states until the program has generated those 
pointers.

Our implementation defines the class as follows:

\[ class := \left\{
    \begin{array}{l l}
    \{0,old frame address\} & \text{For return addresses}\\
    \{1,address\} & \text{For function pointers}
    \end{array} \right.\]

Our system uses the address of the \verb|pointer| as the class for function 
pointers with the highest bit set.  On the x86-64 architecture only 48 bits of 
virtual address space are used, thus leaving us with the 16-bits that we can 
modify safely.  For return pointers we use the old frame pointer as our class.  
Setting the most significant bit for function pointers gives us domain 
separation that ensures an attacker cannot use a function pointer as a return 
address and vice versa.

The 64-bit \verb|pointer| is concatenated with its 64-bit \verb|class| to form 
a 128-bit AES block, which is then encrypted using AES with a 128 bit key.   
Our implementation crucially leverages the AES-NI instructions available on the 
Intel x86-64 architecture~\cite{bib:aesni} to minimize the performance impact 
of these computations.

Including the pointer address in the MAC data ensures that an attacker cannot 
swap a pointer stored in one memory address with a pointer stored in a 
different memory address.  However, the attacker can still replace a pointer 
stored at location $x$ at time~$t$ with a pointer stored at the same location 
$x$ at time $t' < t$.  We refer to this as a {\em replay attack} and discuss 
defenses against it in Section~\ref{sec:replay}.

Our approach generalizes to any hardware supporting fast MACs.
Hardware support for AES, SHA-1, and SHA-256 are available or will be
available on many modern architectures.  This enables efficient
implementation of CCFI on many modern platforms:  ARM 64-bit
architecture includes support for these instructions and
implementations will be widely available for server deployments later
this year~\cite{bib:amdarm}.  The latest SPARC and the PowerPC
architecture include AES instructions~\cite{bib:aesisawiki}.

HMAC-SHA-256 is a good alternative to our AES-based MAC\@.  HMAC may
offer performance benefits because it requires fewer registers for the
key schedule and MAC computation.  At the time of writing the
SHA-1/256 instructions are not yet available on x86 to offer a point
of comparison.

\subsection{Protection Primitives}
\label{sec:primitives}

CCFI protection is built on two main primitives that we provide as compiler 
intrinsics.  This enables our system and an application 
programmer to use these intrinsics to protect any values they deem vulnerable.  
Listing~\ref{lst:builtins} shows the two intrinsics as exposed to the C 
programming language on x86.

\definecolor{lightgrey}{rgb}{0.95,0.95,0.95}
\lstdefinestyle{CStyle}{
    language=C,
    basicstyle=\small\ttfamily,
    frame=tb,
    columns=fullflexible,
    backgroundcolor=\color{lightgrey},
}

\begin{lstlisting}[style=CStyle,label=lst:builtins,caption={Compiler intrinsics 
for protecting and checking a pointer's integrity.}]
__m128d
__builtin_ccfi_macptr(uint64_t ptr,
                      uint64_t class);

uint64_t
__builtin_ccfi_checkptr(uint64_t ptr,
                        uint64_t class);
\end{lstlisting}

The first intrinsic {\tt \_\_builtin\_ccfi\_macptr} it used to protect a 
pointer.  It computes the MAC of the function pointer and the class.  The MAC 
is then stored in a region of memory indexed by the address argument.

The second intrinsic {\tt \_\_builtin\_ccfi\_checkptr} recomputes the MAC and 
compares it against the previously saved MAC\@.  If they are not identical it 
will return a zero value.

These two primitives can be used by a programmer to protect application specific
objects that may affect control flow, like, for example, an index to a jump
table.

\subsection{Stack Protection}

To protect stack-based control flow objects, on each function prologue, we MAC 
both the saved return address and frame pointer (instead of address).  A slot 
on the stack is reserved to store the MAC\@.  In the epilogue, we verify the 
MAC before returning to the caller.  MACing the frame pointer prevents exploits 
that, for example, reposition the upper stack frame to pop and return to 
unexpected return addresses~\cite{nxbit}.

We further optimize leaf functions (functions that do not call out any
other functions) by storing the return address in a register rather
than memory thereby eliminating the need to MAC their return address.

\subsection{Pointer Protection}

Function pointers and vtable pointers are protected in the same way.  They both
point to read-only memory so only the pointer itself needs protection.  On a
store, the MAC is computed, and on a load, the MAC is checked.  Logically, the
MAC should be stored side-by-side with the pointer.  We use a different,
simpler, implementation that does not require resizing function pointers,
structures and classes.  It stores the MAC in an external table, indexed by the
location of the pointer.  A simple hash function can be used to translate a
pointer's address to a slot in the hash table.

C++ adds some complexity in protecting pointers.  Pointers to member functions
of classes are represented by a pair of 64-bit values.  If the first value is
not odd, then it is the address of a non-virtual function.  Otherwise, a virtual
function is being pointed to, and the second 64-bit value indicates the index in
the vtable.

Constructors need to be extended to MAC the vtable pointer as the object is
being instantiated.  Care must be taken to support vtable tables (VTTs), a
condition that arises from multiple inheritance when derived classes share a
common base (C++'s diamond problem).

\subsection{Other Control Flow Protections}

We must also protect other sensitive pointers, specifically the global
offset table (GOT) and global destructors ({\tt .dtors}).  The GOT is
used for dynamic linking and filled by the loader with the addresses
of external library functions.  Global destructors (like global
constructors) are function pointers registered at program load time
and executed at program termination.

To protect these we use an existing mechanism, RELRO~\cite{relro}, which
computes relocations at program load time and marks the GOT and {\tt .dtors}
read-only.  This prevents the attacker from tampering with these sensitive
pointers.

\subsection{Compatibility}

Because our MAC protects the location of a control pointer, each time a control 
pointer is copied, the MAC must be recomputed.  This is not a problem for 
return addresses and vtable pointers because they are not exposed to 
programmers so our system can take care that they are properly handled.  
Function pointers, however, can be copied by the programmer.  When type 
information is preserved, the CCFI compiler can detect that a function pointer 
is being copied and automatically recompute the MAC for the copy.  If, however, 
a function pointer is cast to another type and then copied, the compiler cannot 
detect that a function pointer is being copied so the MAC is not recomputed.  
The result is that if the function pointer is later used, a MAC failure will 
occur.  MAC check will always be present because a function pointer must be 
cast to a function pointer type before being used so the system will err in the 
favor of safety.

In practice, we observed two programs that copy pointers around after casting
them to void, using {\tt memcpy}.  Apache and nginx both have a dynamic array
implementation which stores its elements in a void* memory region, and copies
them using {\tt memcpy} if the array is resized.  This array is used for storing
function pointers.  These cases were easy to detect in practice because the
programs would crash at initialization (due to MAC failures).  

We wrote a simpler static analyzer to help programmers find cases where type
information is lost for function pointers and manual MAC checks may be
necessary.  The Nginx compatibility issue was in fact found by our tool.

\subsection{Replay Attacks} 
\label{sec:replay}

Replay attacks in cryptography exist because an attacker can record a
properly signed message and send it again (without modification) at a
later time.  Protection against replay attacks in cryptography is done
by including a counter or nonce in every message so that the verifier can
detect replayed messages.  

In the context of CCFI, we compute the MAC on a pointer and its
address.  The address functions as a (naive) nonce.  However, it is
still possible to replace the current pointer at location $x$ with an
old pointer previously stored at location $x$ and potentially disrupt
control flow.  We stress that this is the only attack possible 
with our basic MAC method.  This problem does not exist with globals,
because they can exist in only one place, and their address
effectively acts as a unique nonce.  The problem with the heap and stack is
that the nonce (i.e., the address) can be reused.

One approach to fixing potential replays is to add randomness to
every memory allocation (including stack frames) to prevent addresses
from aligning.  This reduces the likelihood that two function pointers
share the same memory address (and hence same nonce),
which prevents replay attacks.  Our final CCFI design MACs the object's
data and address and furthermore randomizes allocations as described in
Section~\ref{sec:random}.  An alternative that does not rely on
randomization is discussed in Section~\ref{sec:discuss}.

Our stack and heap randomization, used to mitigate replay attacks, is quite 
different from randomization used in ASLR\@.  ASLR is applied once at program 
startup to prevent an attacker from predicting the address of memory objects.  
Our randomization is applied on every allocation to prevent two function 
pointers from always being allocated to the same address. Some systems like 
OpenBSD already randomize this way: every call to {\tt malloc} returns a random 
memory chunk.  We similarly randomly pad stack frames to prevent two return 
addresses from always being placed in the same stack address.  This way even if 
the same call graph is executed multiple times, the location of any return 
address and local function pointer will vary.

\section{Implementation}

Our system is built on the Clang/LLVM compiler framework and supports x86\_64,
tested on FreeBSD\@.  The implementation consists of the following major
components:
\begin{itemize}
\item {\bf Memory randomization:} A change in libc's {\tt malloc} to return
random memory chunks, and an LLVM function pass that randomly offsets the
stack pointer on each call.  This further prevents replay attacks.

\item {\bf LLVM Target:} ABI changes to reserve registers to ensure the 
compiler never leaks the key.  Implement stack protection into the target 
specific code.

\item {\bf Compiler Intrinsics:} {\tt macptr} and {\tt checkptr} compiler 
intrinsics are implemented as machine specific code and made available to the C 
language.

\item {\bf Pointer Protection:} This is a high-level LLVM pass that identifies 
critical function pointers and vtable pointers that must be protected.  A small 
runtime library provides error reporting and handling of globals in a central 
place to reduce the increase in code size.

\item {\bf Static analysis tool:}  This finds any possible code that may break
our automatic MAC protection because function pointers are being copied without
type information and hence the MAC is not being automatically recomputed by the
CCFI compiler.  
\end{itemize}

Any application wishing to be hardened with CCFI must be recompiled along with
all of its dependencies.  We provide a command-line compatible wrapper to clang
and clang++.

\subsection{ABI Changes}

We implemented our MAC using the AES-NI instructions on x86.  These instructions
take their arguments in XMM registers.  A 128-bit AES key expands to 11 128-bit
values, requiring 11 XMM registers (each 128-bits wide) to hold the key.

XMM registers are normally used for floating point and vector operations in the 
AMD64 ABI specification.  They are also used for argument passing.  We must 
therefore reserve 11 of these registers to hold our expanded key.  An 
additional scratch XMM register is needed while computing the AES rounds.  This 
register must not be used for argument passing or it would be clobbered during 
our AES computation during the function's prologue.  It can, however, be used 
in the function's body as a temporary.  Table~\ref{tab:abichange} shows how we 
change the ABI we made to manage our AES encryption.

\begin{table}
\centering
\begin{tabular}{|l|l|l|}
\hline
{\bf Registers}    & {\bf SysV ABI} & {\bf CCFI ABI} \\ \hline
{\tt xmm0-xmm3}    & Arguments      & Arguments \\ \hline
{\tt xmm4}         & Arguments      & Temporary* \\ \hline
{\tt xmm5-xmm7}    & Arguments      & Expanded Key \\ \hline
{\tt xmm8-xmm15}   & Temporary      & Expanded Key \\ \hline
\end{tabular}
\caption{Shows how the XMM register usage in the AMD64 SysV ABI differs from 
that of our ABI\@.  The general purpose registers, float point stack and MMX 
registers remain unchanged in the new ABI\@.  The changes only affect the XMM 
registers, which are used for floating point and vector operations. *When 
optimizing protection for leaf functions stack protection uses XMM4 to store 
the return instruction pointer and frame pointer.}
\label{tab:abichange}
\end{table}

The ABI change reduces the available registers for floating point and vector 
computation to only five.  This limits the compiler's ability to keep more 
variables in the registers and thus can have a substantial impact on 
performance.  Code that does not use floating point or vector math operations 
should not notice a performance impact from this change.  Some programs use the 
XMMs also for copying memory, zeroing memory, and similar operations.  These 
tasks typically require only one or two XMM registers.

\subsection{Memory Randomization}
\label{sec:random}

To prevent replay attacks where control pointers align, each memory and stack
frame allocation is randomized.  The basic idea is to add a random offset to
each {\tt malloc} and stack frame.  There is a trade-off between how much
virtual memory to waste and how much entropy to add.  OpenBSD already implements
this for {\tt malloc} and we use their same entropy parameter of four bits.

We implemented randomized allocations in FreeBSD's libc {\tt malloc} by adding a
random offset to each allocated chunk.  Unlike OpenBSD, we are not allowed to
store a randomness source in memory because the attacker can modify this as per
our threat model.  Instead, we are required to use registers.  Ideally we would
use Intel's random instruction~\cite{rdrand}.  This was not
available on our processor so in the interim we chose to use the CPU's cycle
counter.  The attacker would have to both align memory layout and time (at a
cycle granularity) to conduct an attack.

Stack randomization is implemented similarly.  On each function prologue, we
{\tt alloca} a random size which has the effect of padding the stack frame by a
random value.

\subsection{Stack Protection}

Our stack protection mechanism allocates a local variable to store the
MAC of return address and frame pointer.  The prologue of a function
generates the MAC and stores it.  The epilogue must recompute the MAC,
and compare it, and crash the program if the MAC does not match.  In
the event of a bad MAC, it crashes the program by storing zeros in the
return address and (if there is one) frame pointer, which saves a few
instructions and avoids a branch.


The leaf optimization for stack protection stores the return instruction 
pointer and frame pointer in XMM4.  Rather than encrypting it we just verify 
that it has not been modified.  Since leaf functions do not make any calls we 
can safely rely on a register to store the value.

\subsection{Compiler Intrinsics}

The two portable intrinsics {\tt checkptr} and {\tt macptr} are exposed to the 
programmer and to our protection pass.  Their implementation is machine 
specific depending on the availability of cryptographic instructions.  We 
implemented ours using the AES-NI x86 instructions.  These primitives depend on 
our ABI changes and never leak any part of the expanded key on to the stack.

Both primitives are implemented using a large in-memory hash table that stores
the generated MACs.  The hash table approach allows us to avoid the complexity
associated with wide pointers.  Ideally, we would eventually support wide
pointers, but this makes incremental deployment more difficult as one cannot mix
libraries compiled with and without CCFI as types now have different sizes.  
Using a hash table causes additional performance overhead due to the 
computation of the hash function and potential cache misses associated with 
accessing a separate hash table.  We intend to support wide pointers in the 
future to eliminate these issues.

\subsection{Pointer Protection}

The pointer protection is a module pass which does two things.  First, it goes
through each basic block to find loads and stores of function pointers, adding
calls to {\tt checkptr} and {\tt macptr} respectively.  Care must be taken to
recursively walk every structure, array and vector so that nested function
pointers are found.  When a structure is copied, the code results in an LLVM
{\tt memcpy} intrinsic.  These structures may contain function pointers so we 
must verify and recompute the MAC of the function pointer when this occurs.

The second action of the module pass is to go through all globals and identify
any function pointers (even within structures) that are defined.  These must be
MACed before starting the program.  We create constructor functions that add 
MACs to those.  This is not necessary for globally defined C++ objects because 
actual constructor code, where the MAC is set, will be called.

Some systems calls take or return function pointers.  No special handling is 
needed when these pointers reside in registers, as the compiler already checks 
function pointers when they are loaded into registers.  However, some system 
calls, such as \texttt{sigaction}, exchange structures containing function 
pointers. Instead of modifying the kernel, we modified libc to check pointers 
in argument structures and MAC those in return structures.


\subsubsection{Runtime}

The runtime mostly provides common functions to limit binary size
bloat.  We have a constructor function that is executed on program launch to
allocate a memory region for MAC storage (our hash table).  A global MAC helper
function helps reduce the instantiations of the {\tt macptr} intrinsics
inside constructors.  Lastly, we provide a function to call on failure to help
with debugging and identifying whether it might be a program issue (e.g., 
missing MAC on untyped function pointer copy) or attack.

\subsection{Static Analysis Tool}
We wrote a static analysis tool using Clang's static analyzer to find any code
which may circumvent the automatic MACing of function pointers and therefore
cause bogus MAC failures.  It can be invoked as a wrapper to
\texttt{make} to analyze an
entire application.  It detects and flags the following cases:
\begin{itemize}

\item A {\tt memcpy} where both supplied arguments (before casting) are of
type \verb|void*|.  These could be pointers to data structures containing 
function pointers cast to \verb|void*| in another object file.  In our test 
applications, so far we found this to be the only case where we miss function 
pointer copies.

\item Any place where a function pointer is cast to a non-function pointer type
e.g., unsigned long, or \verb|void*|.

\item Any place where a non-function pointer type is cast back to a function
pointer.
\end{itemize}

We also provide users with a utility function, {\tt ccfi\_memcpy}, that can be
useful for debugging MAC failures due to {\tt memcpy}ing untyped function 
pointers around.  Our {\tt ccfi\_memcpy} analyzes the region being copied and
checks if a MAC is associated with any of the elements in the memory region.  
If so, the MAC is recomputed in the new region.  We used this in Nginx and 
libapr for example, where function pointers were being {\tt memcpy}ied without 
type information and the MAC had to be recomputed.

\section{Security Analysis}
\label{sec:security}

\subsection{Address Aliasing}

Our weakest design point is that two different pointers may be stored at the 
same address, and an attacker could swap those pointers if he happens to have 
observed both of them and their corresponding MACs.  Programs themselves may 
write two different pointers to the same variable, but swapping these two is 
valid from the perspective of the control flow graph.  We only need to prevent 
unrelated variables aliasing to the same address.  The randomization of stack 
and heap layout helps to mitigate this by reducing the probability that 
pointers alias to the same address.  Also return addresses and function 
pointers use different MAC schemes making them fundamentally incompatible.

One way future work could strengthen this is to use a slab allocator to store 
objects of the same type in the same virtual address space without any possible 
overlap with other objects.  As pointed out in our ideal model this can over be 
difficult and some types hide the full type information to be able to implement 
this.  Randomization could be used for the remaining types we cannot reason 
about.  Lastly, types that do not contain sensitive pointers may be a shared 
heap.

The return stack may be addressed using a slab allocator in combination with 
segmented stacks.  The goal again is to prevent address space reuse for the 
stack.  Local sensitive pointers, and return pointers will be protected.  One 
caveat though, similar to all fine-grain CFI approaches we know about, this 
technique would still allow an attacker to return to any function that is a 
valid return target for the current function.

Another avenue for improving the defense is to add a hash of the type signature 
of functions into {\it macptr} and {\it checkptr}.  This would prevent two 
pointers of different types from being swapped as the type signature hash would 
be statically defined in the code using and creating pointers.  This doesn't 
prevent pointers with the same type signature and very different meanings from 
being swapped.  C programs often cast function pointer types, e.g.,~when 
arguments are unused, which we can address by verifying and recomputing the MAC 
with the new type signature.

%

%
%

\subsection{Pointer Table Indices}

Program data that can modify control flow is vulnerable, and we do not defend 
against this in any way.  Developers should use our {\it macptr} and {\it 
checkptr} intrinsics to protect indices into function tables and critical 
conditional values (i.e.,\ whether a connection is authenticated or not).

\section{Evaluation}
We evaluate two aspects:
\begin{enumerate}
\item Do applications break?  When copying function pointers, they must be
reMACed.  This will not occur automatically if a function pointer is cast to a
non-function pointer type.

\item What is the overhead of the MAC computations and checks?
\end{enumerate}

All performance benchmarks were conducted on a computer running FreeBSD 10.0 powered by 
dual Intel Xeon E5620 processors running at 2.4~GHz with four cores each. 
The machine had 48~GBs of RAM and an Intel SSD\@.  A second identical machine 
running Ubuntu Linux was directly connected via gigabit Ethernet to launch 
the network benchmarks.

\subsection{Application Compatibility}
We compiled 21 libraries, 5 servers, and SPEC CINT2006 using CCFI\@. Out of 
these, we only had to modify two lines in libapr and a single line in nginx, 
all of which copied function pointers over with {\tt memcpy}, breaking our 
MAC\@.  In both cases, the programs (nginx and Apache using libapr) crashed 
upon initialization due to a null MAC\@.

We ran our static analysis on nginx and it pointed out three dangerous calls to
{\tt memcpy}.  Two were in a variable sized array implementation which would
{\tt memcpy} its elements to a new buffer when resizing.  The third was in a
resolver code.  Sixteen calls to function pointers being cast to void types were
spotted.  All of these were calls to push function pointers into the array
implementation containing the {\tt memcpy}.  This information directly pointed
us to the problematic {\tt memcpy}.  Interestingly, libapr had the same exact
problem.  A custom array implementation was used to store function pointers in
non-function pointer typed memory.

OpenSSL is a heavy user of function pointers and we were able to run it
unmodified.  We had to disable some of the assembly optimizations that used our
reserved XMM registers.  We could have modified the assembly code to save and
restore our key in YMM registers.  In fact, we are looking to implement this
directly in LLVM to automatically support hand written assembly code that uses
our reserved XMMs.

\subsection{Microbenchmarks}
Our system proposes to compute AES on every call, return and indirect branch.
This seems like a high price to pay but the key to making this practical is the
low latency offered by the AES-NI instructions.

Table~\ref{tab:intrinsics} shows the latency in cycles for each of our intrinsic
functions which essentially run AES on a single block.  We see that the MAC
computation and verification is approximately 40 cycles. 

\begin{table}
\centering
\begin{tabular}{|l|r|}
\hline
Operation           & Cycles \\ \hline
macptr intrinsic    & 40     \\ \hline
checkptr intrinsic  & 39     \\ \hline
\end{tabular}
\caption{Shows the cycles for computing or checking a function pointer.  This 
is only the intrinsic and excludes the conditional statements that are inserted 
when {\tt checkptr} fails.}
\label{tab:intrinsics}
\end{table}

Table~\ref{tab:calltime} examines how the MAC computation time effects function 
call and return times in cycles.  This is our worst case performance because the
function does not do any work.  Stack protection adds approximately 63 cycles 
to the function (70-7).  This value is less than twice the average MAC 
computation time because it is the operation latency.  The function prologue's 
latency can be hidden partly by the function's body and epilogue computation.  
The epilogue latency could also be hidden if we schedule useful (but safe) work 
to occur after the MAC verification.  Any function performing significant 
amount of computation will mask our fixed overhead of 70 cycles.

The function pointer call latency is listed in the second row.  We see that 
function pointer protection costs an additional 43 cycles.  When enable both 
this numbers increases to 153 cycles.

Finally, two C++ call benchmarks: a non-virtual method pointer call and a
virtual method pointer call are shown.  Calls made through method pointers are 
more expensive because C++ on x86 lowers them into an if-statement that either 
calls the vtable entry if it is virtual otherwise calls the pointer directly.  
Virtual calls are the most expensive because of the extra vtable access.

Overall, CCFI adds a fixed overhead ranging from 70--164 cycles to function 
calls.  Any function doing significant work will amoritize this fixed latency.  
As a reference point, a single cache miss is 300 cycles on modern machines.  
Larger functions enable the processor to take advantage of instruction 
reordering and speculative execution to hide some of this latency.  These 
processor optimizations explain the non-linearity visible in this table.
We evaluate application benchmarks next to measure the overall effect.

\begin{table}
\centering
\begin{tabular}{|l|r|r|r|}
\hline
Operation	& Baseline  & Ptr Prot.	    & CCFI \\ \hline
Func. call	& 7	        & -		        & 70   \\ \hline
Fptr. call	& 7	        & 50	        & 153  \\ \hline
Mthd. call	& 8	        & 53	        & 156  \\ \hline
Vptr. call	& 17	    & 60	        & 164  \\ \hline
\end{tabular}
\caption{Shows the round-trip function call and return for an empty
function in cycles.  The baseline numbers include no protection using an 
unmodified compiler.  CCFI without stack protection shows the overhead when 
only function pointer protection is enabled.  The CCFI column shows the results 
with stack and function pointer protection enabled.}
\label{tab:calltime}
\end{table}


\subsection{SPEC2006 Benchmarks}


\begin{figure*}
\centering
\includegraphics[width=\textwidth]{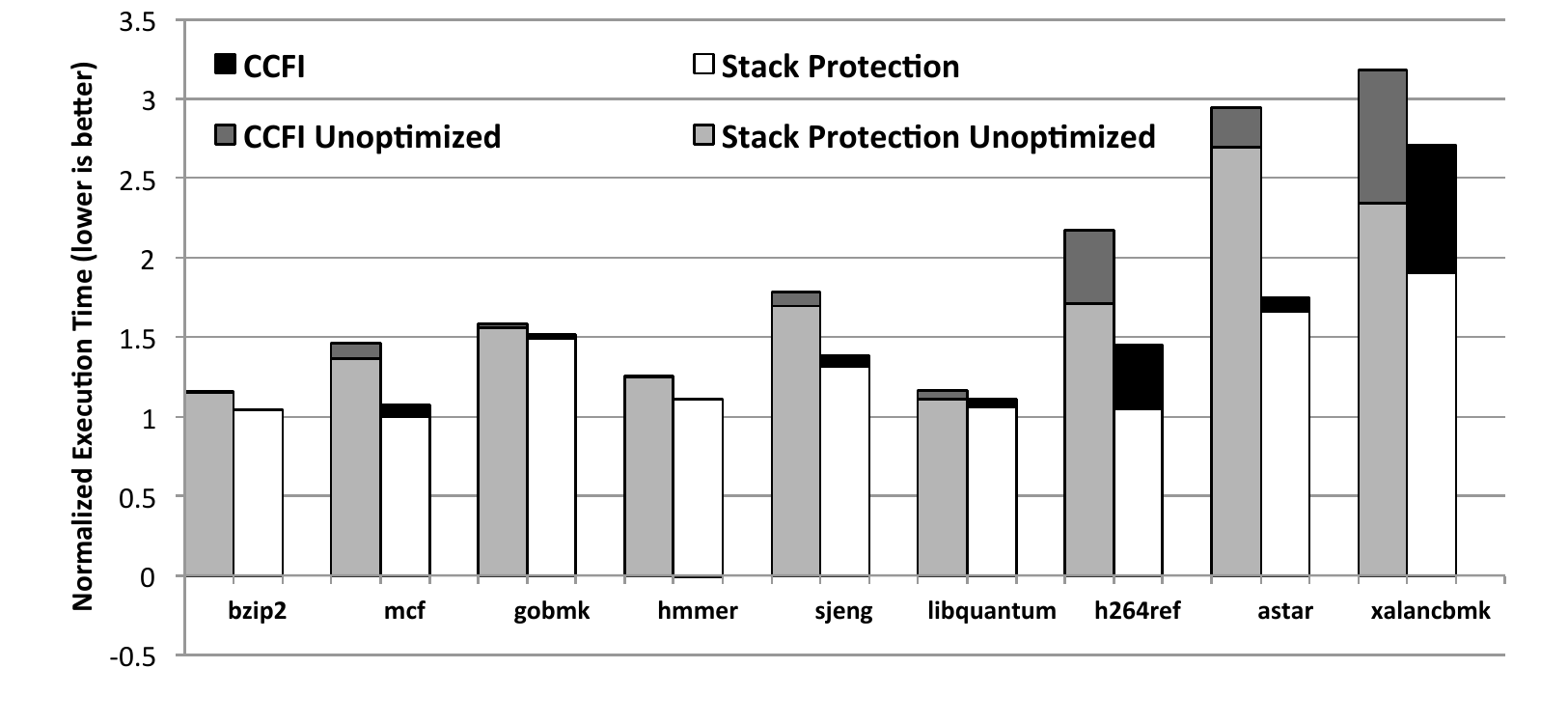}
\caption{Shows the SPEC2006 results.  The left bar is unoptimized CCFI and the
right bar is optimized.  Results are normalized to baseline execution (of 1x).
The bars are stacked to show the overhead of stack protection and function
pointer protection.}
\label{fig:spec}
\end{figure*}

Figure~\ref{fig:spec} shows the results from the SPEC CPU2006 integer
benchmarks.  We omitted the GCC and Perl benchmarks as they crashed when
compiled with the modern vanilla GCC or Clang compilers that we tested.  All
other benchmarks worked both with Clang and CCFI, with no changes to the
benchmark source code.

The Figure uses an unmodified Clang 3.3 compiler as the baseline and the other
results are normalized to this measurement.  We also measured the overhead of
the ABI changes alone which put register pressure, but the result was negligible
so we did not plot it.  All the overhead comes from stack and function pointer
protection.

We show the results of SPEC for full protection with and without the leaf
optimization for stack frames.  The stack protection overhead appears as the
lower half of the bar in each of the two cases.  We measured an average of 45\%
overhead for all benchmarks, and 23\% overhead for the C benchmark.

Function pointer protection overhead becomes more apparent in the C++ 
benchmarks that we have measured.  This is because inheritance depends on
vtable pointers that must be protected.  The C code has very few hot paths 
containing function pointers thus why we do not see a significant performance 
difference between stack protection alone and full protection.

\begin{figure}
\centering
\includegraphics[width=0.5\textwidth]{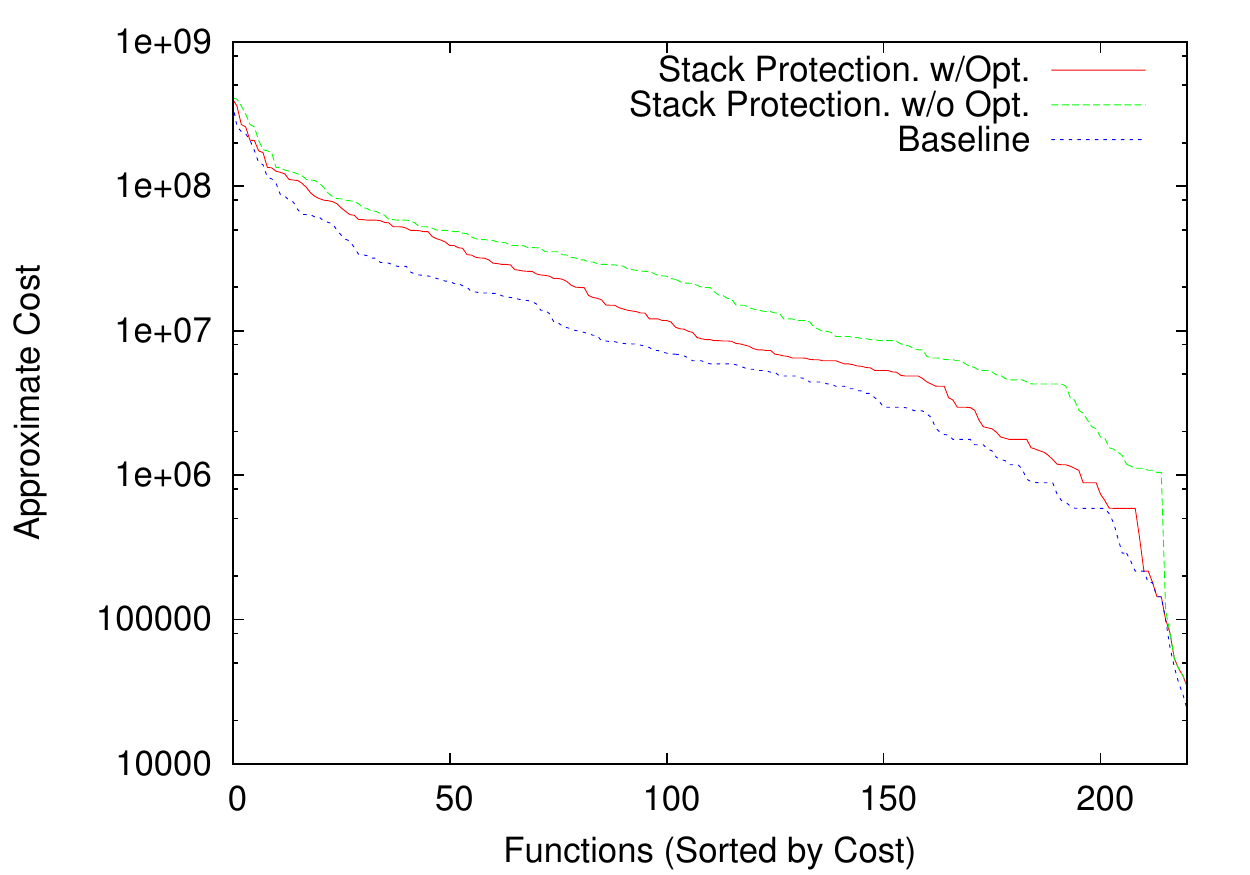}
\caption{Shows the approximate performance cost of the top 220 functions for 
    omnetpp, a C++ benchmark, with vanilla Clang, and stack protection with and 
    without the leaf optimization.  We have many very high frequency smaller 
    sized functions that appear in the graph as the large gap between the vanilla and 
    unoptimized lines.  Our optimization reduces almost half the cost as shown 
    by optimized line which is bounded by the other two.}
\label{fig:omnetpp}
\end{figure}

\begin{figure}
\centering
\includegraphics[width=0.5\textwidth]{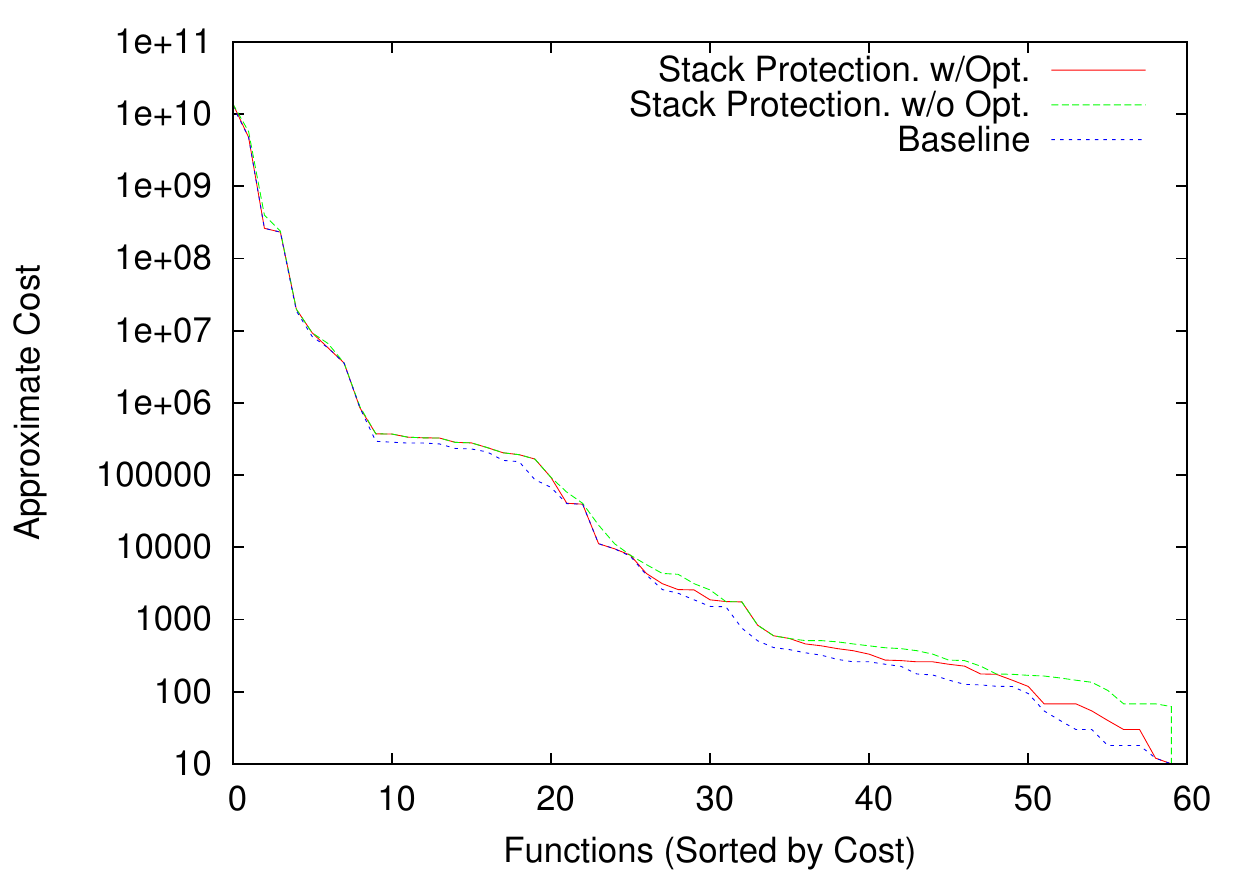}
\caption{Shows the approximate performance cost of the top 60 functions for 
    bzip2, a C benchmark, with vanilla Clang, and stack protection with and 
    without the leaf optimization.  The lines almost completely overlap except
    for the lowest cost functions (far right).  The low cost functions are 
    executed few times and are not very long thus showing more dramatically the 
cost of stack protection.}
\label{fig:bzip2}
\end{figure}

\subsection{Stack leaf optimization gains}
Our stack protection cost dominates in small functions.  The effect worsens when
such functions are called frequently.  To better understand this behavior, we
study our worst and best cases from SPEC (omnetpp and bzip).
Figures~\ref{fig:omnetpp} and~\ref{fig:bzip2} show the approximate total cost
per function (instruction\_count $\times$ number\_of\_calls) for omnetpp and
bzip2, when using different compilers.  Each curve is sorted by function cost.  
The gap between the top curve (stack protection) and the bottom curve (vanilla 
compiler) shows the overhead of stack protection.  The middle curve 
approximates the cost graph with the leaf optimization.  In the omnetpp case, 
there are many frequent calls to smaller functions (typical in C++) and hence 
the higher overhead.  This is indicated by the middle curve that hugs the 
unoptimized curve on the left side of the graph (costly functions).  On the
remainder of the graph however the optimization pays back as it sits between 
the baseline and unoptimized curve.  Our leaf optimization reduces a 5x 
overhead to 3.5x.

C code represented by bzip2 has many calls to larger functions that is 
indicated by the lines overlapping for most of the graph.  The optimization has 
a smaller impact on overall performance as stack protection contributes to a 
smaller percentage of a functions execution time.

More aggressive function inlining with link time optimization (LTO) should 
bring down the cost of C++ for real world uses.  Pointer protection could be 
eliminated for many small accessor functions that are executed very frequently 
by inlining.

\subsection{Applications}
\begin{table}
\centering
\begin{tabular}{|l|r|r|}
\hline
{\bf Configuration}     & {\bf Vanilla} & {\bf CCFI} \\ \hline
Nginx (https)  		& 207           & 128 \\ \hline
Nginx                   & 16482         & 14103 \\ \hline
Lighttpd		& 22714		& 18516 \\ \hline
Apache			& 25305		& 24537 \\ \hline
\end{tabular}
\caption{\label{tab:web}Webserver request throughput.}
\end{table}

We compiled a number of high performance servers and all their dependencies with
CCFI\@.  Table~\ref{tab:web} shows the request rate when comparing a vanilla 
build of the system compared to CCFI\@.  We used default settings for all 
servers and the ApacheBench benchmarking tool.  In the HTTP case, there is a 
3--18\% overhead depending on the server used.

In the HTTPS case, performance drops by 38\% for two reasons.  First, we
disabled some of the assembly code in OpenSSL which used XMM registers 5--15.
Second, all the intensive math C code felt the XMM register pressure.  Although
we disabled OpenSSL's AES-NI implementation, we used FreeBSD's cryptodev kernel
AES-NI implementation for high speed AES\@.  This comes at the cost of a system
call but is amortized for large messages (anything over 128 bytes will break
even).  Further performance improvements would require changes to OpenSSL's
code.  Specifically, all assembly optimizations would have to be enabled, and
for those using many XMM registers, CCFI's XMMs registers would have to be saved
and restored in YMMs.

\begin{table}
\centering
\begin{tabular}{|l|r|r|}
\hline
{\bf Configuration}     & {\bf Vanilla} & {\bf CCFI} \\ \hline
memcached  		& 283403        & 276006 \\ \hline
redis                   & 107527        & 88496 \\ \hline
\end{tabular}
\caption{\label{tab:memcache}Cache server request throughput.}
\end{table}
We also measured the performance of memcached and redis, shown in
Table~\ref{tab:memcache}.  We used the mutilate tool to benchmark memcached, and
redis' own benchmark tool for redis.  The performance degradation is between
3--18\%.

These results are promising for securing network servers where most of the
overhead comes from IO or complex application code.

\section{Discussion}
\label{sec:discuss}

\subsection{Hardware Mechanisms}

Our CCFI work preserves the immutability of objects through cryptography.  
Other implementation approaches include using other hardware mechanisms like 
paging.  For example, a shadow memory can be used for storing function 
pointers.  This memory is made available only prior to indirect call or jump 
instructions.  Switching page tables using a tagged TLB (or invlpg instruction) 
could be one implementation.  More promising is Supervisor Mode Access 
Protection (SMAP), an upcoming feature that prevents kernels from accessing 
user space memory.  An efficient instruction is available to toggle when a 
kernel is able to read user memory.  This feature could be used in a system 
like Dune~\cite{dune} by running a process in privileged mode, but storing 
function pointers in user space memory.  Before a call, the SMAP bit can be 
flipped to access user space memory, and the bit can then be toggled back in a 
function prologue.

Another way to defend the return stack is to use hardware performance counters
to verify the integrity of the return stack.  This by itself has been done in
previous work~\cite{bib:kbouncer} to have the operating system kernel check the
last few stack frames.  The technique could be applied to user level, but at an
increased cost to the function epilogue.  Another approach would be to store a
few return addresses in registers.  When the registers reserved are all full we
would batch MAC and store these values on the stack.  This optimization allows
us to exploit parallelism available in the AES-NI implementation.  We could see
as much as a four time reduction in stack protection costs.
This technique could help us defend against more sophisticated stack smashing
attacks without the need for randomization.  Potentially by doing a more
expensive MAC computation.

Most existing CFI systems operate on binaries by disassembling them and
instrumenting them.  Our approach was to modify the compiler and so we require
sources or a recompilation of binaries.  We could implement CCFI for unmodified 
binaries using a system like Pin~\cite{pin}.

\subsection{Avoiding Address Space Reuse}

Eliminating the use per-frame stack and per-object heap randomization can offer 
stronger security guarantees.  A simple fix to this is to never reallocate 
address space to different types of objects or stack frames.  This will 
guarantee that two different sensitive pointers will never overlap in the same 
address space.  For the heap we will take any object that contains a sensitive 
pointer and give it a unique pool based on its type to allocate from.  The 
stack on the other hand would require segmented stacks that is supported by 
LLVM to ensure that stack segments belonging to different functions never reuse 
address space.  With no address space reuse our MACs would always be unique and 
give us a stronger guarantee than any CFI system alone.

The heap change may even gain us performance from better cache utilization, 
which is one goal of slab and pool allocators~\cite{bib:slaballoc}.  The stack 
change will cost performance as an allocation is required per-frame, but we can 
benefit from existing work on stack segmentation as it is used in many 
languages today.


\subsection{Just-in-Time Compilers}

Our threat model lets an attacker write to any memory, but assumes that
executable pages are read-only.  Modern JITs often require that pages are 
marked writable and executable at the same time.  One approach is to implement 
the JIT as an external process.  The JIT agent will share a region of memory 
with the main process as read-only and executable.  The agent can then emit 
code and generate valid MACs to pointers to be used by the JIT runtime.

This design hopes to prevent attackers from using the JIT to jump to arbitrary 
code fragments or write into executable memory.  Only allowing the JIT to 
create pointers with valid MACs allows us to omit any oracle functions inside 
the main process.  The JIT may never reuse an entry point as it cannot revoke 
the MAC once generated, but we can reuse memory with the exception the entry 
point byte that would be replaced with an invalid instruction.

\section{Related Work}
\label{sec:related}

Modern operating systems in conjunction with compilers implement several 
security features.  Address Space Layout randomization~\cite{bib:aslr} attempts 
to randomize a program and libraries location in memory to make stack smashing 
attacks against known binaries difficult.  In addition most popular compilers 
including support for stack cookies that attempt to detect stack smashing 
attacks~\cite{stackguard}.  These systems both require recompilation of 
software and have been circumvented by attackers for years in 32-bit systems.  
The BROP attack showed that a generalized attack was practical even on 64-bit 
systems without knowledge of the binary.  While these solutions raise the 
attack's complexity it offers no principled security.

After the initial CFI implementation~\cite{cfi} was introduced by Abadi et al.\ 
there are now many CFI systems built on static analysis techniques to achieve 
security. All these systems classify pointers into several categories such as 
call-sites and function pointers.  Arguably the most secure of these is 
CCFIR~\cite{bib:ccfir} that only classifies pointers into four categories.  
This along with the difficulty of achieving compatibility within the limits of 
static analysis has lead to practical attacks on all known CFI 
systems~\cite{out_of_control,stitchgadgets}.  Cryptographic CFI offers the 
first new approach to CFI since the original paper.  Unlike existing CFI 
systems we require binaries and libraries to be recompiled, as existing 
libraries may leak our key or destroy it.

Forward edge CFI is a new system that enforces fine-grain classification for 
function pointer only~\cite{forwardcfi}.  Like CCFI, forward edge CFI offers 
fine grain classification of function pointers, but does nothing for return 
addresses.  Return addresses are left unprotected falling back to existing 
mechanisms that are known to be weak.  As with previous CFI systems, forward 
edge CFI does not have the runtime benefits that CCFI does.

Another very related work is PointGuard~\cite{bib:pointguard}.  The PointGuard 
system exclusive-or's all function pointers with a random value chosen at 
startup.  In a way this can be thought of like pointer encryption except it 
assumes that attackers will only read or modify a single pointer.  Once an 
attacker has read several pointers the secret exclusive-or value can be 
computed.  Cryptographically secure encryption (or MAC'ing) by itself provides 
little as functions can be swapped.  CCFI improvement over PointGuard is 
realizing the connection between inputs to a MAC and CFI.  Another problem is 
that modifying pointers in-place meant that a lot more program/library changes 
are required.  Pointers had to be manually decrypted/encrypted when issuing 
system calls.

Several systems use memory protection hardware to protect the return stack such 
as shadow stack.  The StackGhost system relied on register windows and OS 
support on the SPARC architecture to provide stack smashing 
protection~\cite{bib:stackghost}.  StackShield implemented a shadow stack using 
the data segment so that it would not be susceptible to stack smashing 
attacks~\cite{stackshield}.  These systems do not protect local function 
pointers stored on the stack.  Some shadow stack implementations on x86 use 
segmentation to isolate the shadow stack, such that an attacker could not 
overwrite it without the use of a special instruction prefix.  This CPU feature 
is not supported by any popular architecture today including x86-64 and thus an 
attacker with a stronger threat model could attack the shadow stack.  

Another hardware based approach kBouncer~\cite{bib:kbouncer} used performance 
counters on x86 to record the last few stack entries, and have the operating 
system verify it during the execution of a system call.  This technique only 
protects the top few levels (2--16 depending on the processor support) of the 
stack.

\section{Conclusion}

We showed that cryptographic control flow integrity is a viable
approach to protecting program control flow on modern processors.
Our system ensures that an attacker who has random read
access to memory cannot tamper with control flow data, such as 
return addresses and function pointers, without being detected.
While attackers can cause the program to crash, they cannot alter
control flow to execute code of their choice.

Our implementation of CCFI classifies pointers by the address at which they are 
stored and a single bit to differentiate return pointers from function 
functions.  Classifying pointers by a runtime attribute (addresses) was not 
previously possible.  With no static analysis our modified Clang/LLVM compiler 
can build protected binaries with fine grain control flow integrity.

Our implementation is general enough to integrate with fine grain CFI, when a 
practical CFI analysis is made available in Clang.  This can work in 
conjunction with our address based classification to restrict control flow 
further than any compile-time solution can.

We experimented with our CCFI system on a number of large software packages.  
In all cases the packages compiled with no problems after changing at most one 
line of code in each package.

Clearly a cryptographic system that provides strong control flow
protection must incur some performance cost.  By optimizing the system
and leveraging hardware support for AES available in modern processors
we were able to achieve between 3--18\% slowdown over the unprotected system.  In
many environments this is a worthwhile trade off given the strong
protection it provides.  

This work shows how to protect control flow structures, but does not
protect other data in memory.  It would be interesting to explore
extending the cryptographic protection described in this work to
protect other memory structures, including structures holding
application data.  This will potentially prevent attacks that exploit
data flow vulnerabilities such as Heartbleed~\cite{heartbleed}.  We
leave this as an interesting direction for future work.

\bibliographystyle{unsrt}
\bibliography{paper}

\end{document}